\documentclass[aps,preprint,amsmath,amssymb]{revtex4}

\usepackage{graphicx}
\begin{document}

\title{Electromagnetic properties of the neutrinos in $\gamma p$ collision at the LHC}

\author{\.{I}. \c{S}ahin}
\email[]{inancsahin@karaelmas.edu.tr}
 \affiliation{Department of Physics, Zonguldak Karaelmas University, 67100 Zonguldak, Turkey}

\begin{abstract}
We have investigated non-standard $\nu \bar \nu \gamma $ and $\nu
\bar \nu \gamma \gamma$ couplings via $\nu \bar \nu$ production in a
$\gamma p$ collision at the LHC. We obtain 95\% confidence level
bounds on $\nu \bar \nu \gamma $ and $\nu \bar \nu \gamma \gamma$
couplings by considering three different forward detector
acceptances; $0.0015 <\xi < 0.15$, $0.0015 <\xi < 0.5$ and $0.1 <\xi
< 0.5$. We show that the reaction $p p\to p\gamma p\to p \nu \bar
\nu q X$ provides more than eight orders of magnitude improvement in
neutrino-two photon couplings compared to LEP limits.
\end{abstract}
\pacs{13.15+g,14.60.St,12.60.-i}

\maketitle

\section{Introduction}
Neutrinos play an important role in the evolution of the universe
and in many astrophysical processes. Their properties are under
intense study and some rigorous restrictions have been obtained on
their anomalous properties from controlled experiments and
astrophysical observations. The Large Hadron Collider (LHC) offer
the opportunity of a very rich physics program. It is remarkable to
examine the potential of LHC for probing anomalous neutrino
properties.

In the minimal extension of the standard model (SM) with massive
neutrinos, radiative corrections induce tiny couplings of $\nu \bar
\nu \gamma $ and $\nu \bar \nu \gamma \gamma$
\cite{Schrock,Marciano,Lynn,Crewther,Feinberg}. Despite the fact
that minimal extension of the SM induces very small couplings, there
are several models beyond the SM predicting relatively large $\nu
\bar \nu \gamma $ and $\nu \bar \nu \gamma \gamma$ couplings.
Therefore it is meaningful to search for electromagnetic properties
of the neutrinos in a model independent way. Electromagnetic
properties of the neutrinos have important implications on particle
physics, astrophysics and cosmology.
Probing electromagnetic structure of the neutrinos is important for
understanding the physics beyond the SM and contributes to the
studies in astrophysics and cosmology.

ATLAS and CMS collaborations have a program of forward physics with
extra detectors located at distances of 220m and 420m from the
interaction point \cite{royon,albrow,avati}. Physics program of this
new instrumentation covers soft and hard diffraction, high energy
photon-induced interactions, low-x dynamics with forward jet
studies, large rapidity gaps between forward jets, and luminosity
monitoring
\cite{fd1,fd2,fd3,fd4,fd5,fd6,fd7,fd8,fd9,fd10,fd11,fd12,fd13,fd14,fd15,fd16,fd17,fd18,rouby}.
These forward detector equipment allows to detect intact scattered
protons with some momentum fraction loss
$\xi=(|\vec{p}|-|\vec{p}^{\,\,\prime}|)/|\vec{p}|$. Here $\vec{p}$
is the momentum of incoming proton and $\vec{p}^{\,\,\prime}$ is the
momentum of intact scattered proton. Complementary to proton-proton
interactions, forward detector equipment at the LHC allows to study
photon-photon and photon-proton interactions at energies higher than
at any existing collider. Photon induced reactions in a
hadron-hadron collision were observed in the measurements of CDF
collaboration \cite{cdf1,cdf2,cdf3,cdf4}. The reactions such as $p
\bar p\to p \gamma \gamma \bar p\to p e^+ e^- \bar p$
\cite{cdf1,cdf4}, $p \bar p\to p \gamma \gamma \bar p\to p\; \mu^+
\mu^- \bar p$ \cite{cdf3,cdf4}, $p \bar p\to p \gamma \bar p\to p\;
J/\psi\;(\psi(2S)) \bar p$ \cite{cdf3} were verified experimentally.
These results raise interest on the potential of LHC as a
photon-photon and photon-proton collider.

\begin{figure}
\includegraphics[scale=1]{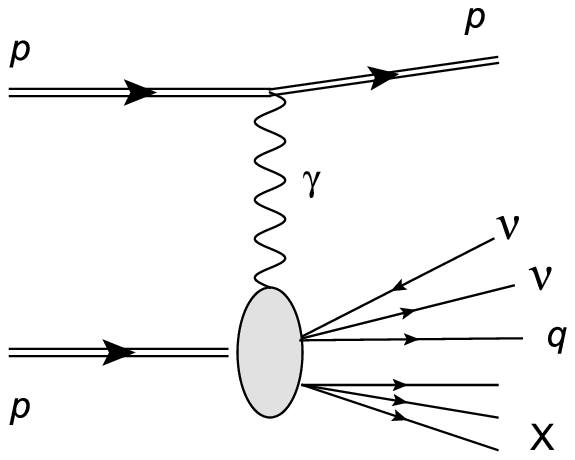}
\caption{Schematic diagram for the reaction $p p\to p\gamma p\to p
\nu \bar \nu q X$. \label{fig1}}
\end{figure}

In this paper we have investigated anomalous $\nu \bar \nu \gamma $
and $\nu \bar \nu \gamma \gamma$ couplings via $\nu \bar \nu$
production in a $\gamma p$ collision at the LHC. A quasi-real photon
emitted from one proton beam can interact with the other proton and
produce $\nu \bar \nu$ pair through deep inelastic scattering.
Emitted quasi-real photons are described by equivalent photon
approximation (EPA) \cite{budnev,Baur,fd18}. Their virtuality is
very low and it is a good approximation to assume that they are
on-mass-shell. In Fig.\ref{fig1} we show a schematic diagram for our
main reaction $p p\to p\gamma p\to p \nu \bar \nu q X$. Intact
scattered protons after the collision can be detected by the forward
detectors. In correlation with central detectors a distinctive
signal for the $\gamma p$ collision can be identified. In
particular, two experimental signatures arise for any reaction in a
$\gamma p$ collision \cite{rouby}:  In the framework of EPA, emitted
quasi-real photons from the protons have a low virtuality and
scattered with small angles from the beam pipe. Therefore when a
proton emits a quasi-real photon it should also be scattered with a
small angle. Hence, intact scattered protons exit the central
detector without being detected. This causes a decrease in the
energy deposit in the corresponding forward region compare to the
case which the proton remnants are detected by the calorimeters.
Consequently, for a $\gamma p$ collision, one of the forward regions
of the central detector  has a significant lack of energy. The
region which is lack of particles defines a forward large rapidity
gap. Usual $pp$ deep inelastic processes can be rejected by applying
a selection cut on this quantity. Another experimental signature is
provided by the forward detectors. When an intact proton is
scattered with a large pseudorapidity it escape detection from the
central detectors. But since its energy is lower than the beam
energy, its trajectory decouples from the beam path into the very
forward region. Forward detectors can detect particles with a large
pseudorapidity. The detection of intact protons by the forward
detectors provides a characteristic signature. The detection of the
$p p\to p\gamma p\to p \nu \bar \nu q X$ reaction involves in
addition to the above signatures, a missing energy signature due to
$\nu \bar \nu$ production. Missing energy signature due to neutrinos
has a different characteristic compare to the case of missing
protons. First, a missing energy signature due to neutrinos is not
only observed in the forward regions of the central detector but
also central regions of it. Second, in the case of missing protons
there should be a missing charge equivalent to the charge of proton.
Third and the most important difference is that protons scattered
with small angles are not genuinely missing. Forward detectors can
detect them. On the other hand, this is not the case for neutrinos.

Forward detectors have a capability to detect intact scattered
protons with momentum fraction loss in the interval $\xi_{min} <\xi
< \xi_{max}$. The interval $\xi_{min} <\xi < \xi_{max}$ is called
the acceptance of the forward detectors. The acceptance of $0.0015
<\xi < 0.15$ is proposed by the ATLAS Forward Physics (AFP)
Collaboration \cite{royon,albrow}. On the other hand, CMS-TOTEM
forward detector scenario propose an acceptance of $0.0015 <\xi <
0.5$ \cite{avati}. Since the forward detectors can detect protons in
a continuous range of $\xi$ one can impose some cuts and choose to
work in a subinterval of the whole acceptance region. In this paper
we consider three different forward detector acceptances; $0.0015
<\xi < 0.15$, $0.0015 <\xi < 0.5$ and $0.1 <\xi < 0.5$.

New physics probes in photon-induced reactions at the LHC have been
discussed in the literature
\cite{fd11,fd12,fd13,lhc1,lhc2,lhc3,lhc4,lhc5,lhc6,lhc7,lhc8,lhc9,lhc10}.
It was shown in Ref.\cite{lhc10} that $\gamma \gamma$ collision at
the LHC has a great potential to probe $\nu \bar \nu \gamma \gamma$
couplings. It provides more than seven orders of magnitude
improvement in $\nu \bar \nu \gamma \gamma$ couplings compared to
LEP limits \cite{lhc10}.

\section{Effective lagrangian and cross sections}
We employ the following effective lagrangian for non-standard
$\nu\bar{\nu}\gamma$ coupling \cite{Larios1,Maya,Bell,Larios2}
\begin{eqnarray}
\label{nunuphoton} {\cal
L}=\frac{1}{2}\mu_{ij}\bar{\nu}_{i}\sigma_{\mu\nu}\nu_{j}F^{\mu\nu}
\end{eqnarray}
where $\mu_{ii}$ is the magnetic moment of $\nu_i$ and $\mu_{ij}$
$(i\neq j)$ is the transition magnetic moment. In the above
effective lagrangian new physics energy scale $\Lambda$ is absorbed
in the definition of $\mu_{ij}$. Non-standard
$\nu\bar{\nu}\gamma\gamma$ coupling can be described by the
following dimension 7 effective
lagrangian\cite{Nieves,Ghosh,Feinberg,Liu,Gninenko,Larios2}
\begin{eqnarray}
\label{nunuphotonphoton} {\cal
L}=\frac{1}{4\Lambda^3}\bar{\nu}_{i}\left(\alpha^{ij}_{R1} P_R+
\alpha^{ij}_{L1} P_L\right)\nu_{j}\tilde
{F}_{\mu\nu}F^{\mu\nu}+\frac{1}{4\Lambda^3}\bar{\nu}_{i}\left(\alpha^{ij}_{R2}
P_R+ \alpha^{ij}_{L2} P_L\right)\nu_{j} F_{\mu\nu}F^{\mu\nu}
\end{eqnarray}
where $P_{L(R)}=\frac{1}{2}(1\mp\gamma_5)$, $\tilde
{F}_{\mu\nu}=\frac{1}{2}\epsilon_{\mu\nu\alpha\beta}F^{\alpha\beta}$,
$\alpha^{ij}_{Lk}$ and $\alpha^{ij}_{Rk}$ are dimensionless coupling
constants. We will consider Dirac neutrino case and obtain model
independent bounds on couplings in the effective lagrangians
(\ref{nunuphoton}) and (\ref{nunuphotonphoton}).

We have rigorous experimental bounds on neutrino magnetic moment
obtained from neutrino-electron scattering experiments with reactor
neutrinos. These are at the order of $10^{-11}\mu_B$
\cite{Li,Wong1,Wong2,Daraktchieva}. Bounds derived from solar
neutrinos are at the same order of magnitude \cite{Arpesella}. On
the other hand we have more restrictive bounds obtained from
astrophysical observations. For instance, bounds derived from energy
loss of astrophysical objects give about an order of magnitude more
restrictive bounds than reactor and solar neutrino probes
\cite{Raffelt,Castellani,Catelan,Ayala,Barbieri,Lattimer,Heger}.
Although there is a great amount of work on $\nu \bar \nu \gamma $
coupling, $\nu \bar \nu \gamma \gamma$ coupling has been much less
studied in the literature. Current experimental bounds on this
coupling are obtained from rare decay $Z\to \nu \bar{\nu}
\gamma\gamma$ \cite{Larios2} and the analysis of $\nu_\mu N\to \nu_s
N$ conversion \cite{Gninenko}. The following upper bound has been
obtained from the LEP data on $Z\to \nu \bar{\nu} \gamma\gamma$
decay \cite{Larios2}
\begin{eqnarray}
\label{leplimit} \left[\frac{1 GeV}{\Lambda}\right]^6
\sum_{i,j,k}\left(|\alpha^{ij}_{Rk}|^2+|\alpha^{ij}_{Lk}|^2\right)\leq2.85\times10^{-9}
\end{eqnarray}
The analysis of the process $\nu_\mu N\to \nu_s N$ conversion via
Primakoff effect in the external Coulomb field of the nucleus $N$
yields about two orders of magnitude more restrictive bound than
$Z\to \nu \bar{\nu} \gamma\gamma$ decay \cite{Gninenko}.

\begin{figure}
\includegraphics[scale=1]{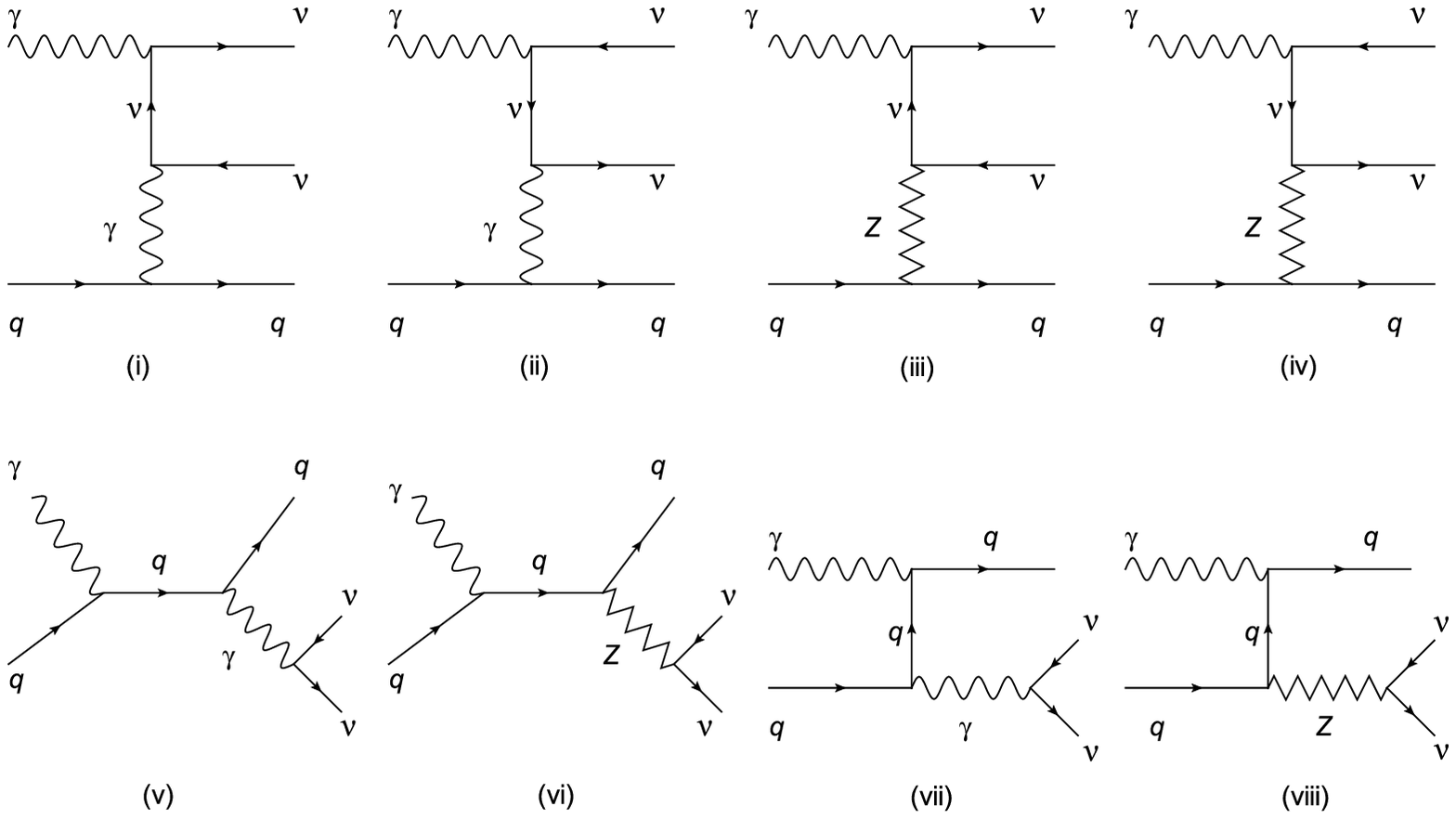}
\caption{Tree-level Feynman diagrams for the subprocess $\gamma q
\to \nu \bar \nu q$ ($q=u,d,c,s,b,\bar u,\bar d, \bar c, \bar s,
\bar b$) in the presence of non-standard $\nu \bar \nu \gamma$
coupling.\label{fig2}}
\end{figure}

We consider the following  subprocesses of our main reaction $p p\to
p\gamma p\to p \nu \bar \nu q X$
\begin{eqnarray}
\label{subprocesses} &&\text{(i)}\;\;\gamma u \to \nu \bar \nu u
\;\;\;\;\;\;\;\;\;\;\;\;\text{(vi)}\;\;\gamma \bar u \to \nu \bar
\nu \bar u \nonumber
\\ &&\text{(ii)}\;\;\gamma d \to \nu \bar \nu d
\;\;\;\;\;\;\;\;\;\;\;\;\text{(vii)}\;\;\gamma \bar d \to \nu \bar
\nu \bar d\nonumber
\\&&\text{(iii)}\;\;\gamma c \to \nu \bar \nu c
\;\;\;\;\;\;\;\;\;\;\text{(viii)}\;\;\gamma \bar c \to \nu \bar \nu \bar c \\
&&\text{(iv)}\;\;\gamma s \to \nu \bar \nu s
\;\;\;\;\;\;\;\;\;\;\;\text{(ix)}\;\;\gamma \bar s \to \nu \bar \nu
\bar s \nonumber \\&&\text{(v)}\;\;\gamma b \to \nu \bar \nu b
\;\;\;\;\;\;\;\;\;\;\;\;\;\text{(x)}\;\;\gamma \bar b \to \nu \bar
\nu \bar b \nonumber
\end{eqnarray}
In the presence of the effective interaction (\ref{nunuphoton}) each
of the subprocesses is described by eight tree-level diagrams
(Fig.\ref{fig2}). As we see from Fig.\ref{fig2}, six diagram
contains non-standard $\nu \bar \nu \gamma$ interaction. The
analytical expression for the polarization summed amplitude square
is quite lengthy so we do not present it here. But it can be written
in the following form:
\begin{eqnarray}
\label{amplitude1}
\sum_{i,j}\langle|M|^2\rangle=\left(\sum_{i,j,m,n}\mu_{im}\mu_{mj}\mu_{in}\mu_{nj}\right)F_1
+\left(\sum_{i,j}\mu_{ij}^2\right)F_2+\left(\sum_{i,m}\mu_{im}\mu_{mi}\right)F_3+F_4
\end{eqnarray}
where
\begin{eqnarray}
\label{F1} F_1=&&|M_1|^2+|M_2|^2+M_1^\dagger M_2+M_2^\dagger
M_1\nonumber \\
F_2=&&\nonumber|M_3|^2+|M_4|^2+|M_5|^2+|M_7|^2+M_3^\dagger
M_5+M_5^\dagger M_3+M_3^\dagger M_7+M_7^\dagger M_3\nonumber \\
&&+M_4^\dagger M_5+M_5^\dagger M_4+M_4^\dagger M_7+M_7^\dagger
M_4+M_5^\dagger M_7+M_7^\dagger M_5\nonumber \\
F_3=&&M_1^\dagger M_6+M_6^\dagger M_1+M_1^\dagger M_8+M_8^\dagger
M_1+M_2^\dagger M_6+M_6^\dagger M_2+M_2^\dagger M_8+M_8^\dagger
M_2\nonumber \\
F_4=&&3\left(|M_6|^2+|M_8|^2+M_6^\dagger M_8+M_8^\dagger
M_6\right)\nonumber
\end{eqnarray}
Here, $M_i\,(i=1,..,8)$ are the "flavor independent" amplitudes of
the diagrams shown in Fig.\ref{fig2}. By using a phrase "flavor
independent" we mean the part of the amplitude which does not
contain the coupling constants $\mu_{ij}$ or equivalently amplitude
with $\mu_{ij}$=1. Since the structure of the effective lagrangian
(\ref{nunuphoton}) is same for each neutrino flavor and we omit the
mass of neutrinos, $M_i$ amplitudes are independent of the neutrino
flavor. In the definitions of $F_i$ functions, average over initial
spins and sum over final spins is implied. Since it is impossible to
discern final neutrino flavor, in Eqn.(\ref{amplitude1}) we perform
a sum over flavor indices $i$ and $j$.

In the case of effective interaction (\ref{nunuphotonphoton}) we
have three tree-level diagrams (Fig.\ref{fig3}). The polarization
summed amplitude square can be written in the form:
\begin{eqnarray}
\label{amplitude2}
\sum_{i,j}\langle|M|^2\rangle=(\alpha_1^2+\alpha_2^2)G_1+G_2
\end{eqnarray}
where
\begin{eqnarray}
\label{alpha1}
\alpha_1^2=\sum_{i,j}\left[|\alpha^{ij}_{R1}|^2+|\alpha^{ij}_{L1}|^2\right]
,\;\;\;\;\;\;\;\alpha_2^2=\sum_{i,j}\left[|\alpha^{ij}_{R2}|^2+|\alpha^{ij}_{L2}|^2\right]
\end{eqnarray}
and $G_1, G_2$ are some functions of the momenta. Again a sum over
indices $i$ and $j$ has been performed. Analytical expressions for
$G_1$ and $G_2$ are given in the Appendix.

In the framework of EPA, equivalent photon spectrum of virtuality
$Q^2$ and energy $E_\gamma$ is given by the following formula
\cite{budnev,Baur,fd18}
\begin{eqnarray}
\frac{dN_\gamma}{dE_{\gamma}dQ^{2}}=\frac{\alpha}{\pi}\frac{1}{E_{\gamma}Q^{2}}
[(1-\frac{E_{\gamma}}{E})
(1-\frac{Q^{2}_{min}}{Q^{2}})F_{E}+\frac{E^{2}_{\gamma}}{2E^{2}}F_{M}]
\end{eqnarray}
where
\begin{eqnarray}
&&Q^{2}_{min}=\frac{m^{2}_{p}E^{2}_{\gamma}}{E(E-E_{\gamma})},
\;\;\;\; F_{E}=\frac{4m^{2}_{p}G^{2}_{E}+Q^{2}G^{2}_{M}}
{4m^{2}_{p}+Q^{2}} \\
G^{2}_{E}=&&\frac{G^{2}_{M}}{\mu^{2}_{p}}=(1+\frac{Q^{2}}{Q^{2}_{0}})^{-4},
\;\;\; F_{M}=G^{2}_{M}, \;\;\; Q^{2}_{0}=0.71 \mbox{GeV}^{2}
\end{eqnarray}
Here E is the energy of the incoming proton beam and $m_{p}$ is the
mass of the proton. The magnetic moment of the proton is
$\mu^{2}_{p}=7.78$. $F_{E}$ and $F_{M}$ are functions of the
electric and magnetic form factors. In the above EPA formula,
electromagnetic form factors of the proton have been taken into
consideration. After integration over $Q^2$ in the interval
$Q^{2}_{min}-Q^{2}_{max}$, equivalent photon spectrum can be written
as \cite{fd11}
\begin{eqnarray}
\label{spectrum} \frac{dN_\gamma}{dE_{\gamma}}=\frac{\alpha}{\pi
E_{\gamma}}
\left(1-\frac{E_{\gamma}}{E}\right)\left[\varphi\left(\frac{Q^{2}_{max}}{Q^{2}_0}\right)
-\varphi\left(\frac{Q^{2}_{min}}{Q^{2}_0}\right)\right]
\end{eqnarray}
where the function $\varphi$ is defined by
\begin{eqnarray}
\varphi(x)=(1+ay)\left[-ln(1+\frac{1}{x})+\sum_{k=1}^3\frac{1}{k(1+x)^k}\right]+\frac{y(1-b)}{4x(1+x)^3}\nonumber\\
+c\left(1+\frac{y}{4}\right)\left[ln\left(\frac{1-b+x}{1+x}\right)+\sum_{k=1}^3\frac{b^k}{k(1+x)^k}\right]
\end{eqnarray}
where
\begin{eqnarray}
y=\frac{E_\gamma^2}{E(E-E_\gamma)},\;\;\;\;\;\;a=\frac{1+\mu^{2}_{p}}{4}+\frac{4m^2_{p}}{Q^{2}_0}\approx7.16 \nonumber\\
b=1-\frac{4m^2_{p}}{Q^{2}_0}\approx-3.96,\;\;\;\;\;\;c=\frac{\mu^{2}_{p}-1}{b^4}\approx0.028
\end{eqnarray}
The contribution to the integral above $Q^{2}_{max}\approx2\;GeV^2$
is negligible. Therefore during calculations we set
$Q^{2}_{max}=2\;GeV^2$.

The cross section for the complete process $p p\to p\gamma p\to p
\nu \bar \nu q X$ can be obtained by integrating the cross section
for the subprocess $\gamma q \to \nu \bar \nu q$  over the photon
and quark spectra:
\begin{eqnarray}
\sigma\left(p p\to p\gamma p\to p \nu \bar \nu q
X\right)=\int_{x_{1\; min}}^{x_{1\;max}} {dx_1 }\int_{0}^{1} {dx_2}
\left(\frac{dN_\gamma}{dx_1}\right)\left(\frac{dN_q}{dx_2}\right)\hat{\sigma}_{\gamma
q \to \nu \bar \nu q}(\hat s)
\end{eqnarray}
Here, $x_1=\frac{E_\gamma}{E}$ and $x_2$ is the momentum fraction of
the proton's momentum carried by the quark. At high energies greater
than proton mass it is a good approximation to write
$\xi=\frac{E_\gamma}{E}$. Therefore upper and lower limits of the
$dx_1$ integral are $x_{1\;max}=\xi_{max}$ and
$x_{1\;min}=\xi_{min}$. $\frac{dN_q}{dx_2}$ is the quark
distribution function of the proton. The virtuality of the quark is
taken to be ${Q^\prime}^2={m_Z}^2$ during calculations. In our
calculations parton distribution functions of Martin, Stirling,
Thorne and Watt \cite{pdf} have been used. We always sum all the
contributions from subprocesses given in Eq.(\ref{subprocesses}).

Assuming only one of the matrix element $\mu_{ij}$ is different from
zero, say $\mu$, we plot the integrated total cross section of the
process $p p\to p\gamma p\to p \nu \bar \nu q X$ as a function of
anomalous coupling $\mu$ in Fig.\ref{fig4}. We see from
Fig.\ref{fig4} that cross sections for the acceptances $0.0015 <\xi
< 0.15$ and $0.0015 <\xi < 0.5$ are close to each other. On the
other hand, cross section for $0.1 <\xi < 0.5$ is considerably
smaller than others. In Fig.\ref{fig5} we plot the integrated total
cross section of the process $p p\to p\gamma p\to p \nu \bar \nu q
X$ as a function of anomalous coupling $\alpha_1$ for the
acceptances $0.0015 <\xi < 0.15$, $0.0015 <\xi < 0.5$ and $0.1 <\xi
< 0.5$. Cross sections in Fig.\ref{fig5} reflect similar behavior to
Fig.\ref{fig4}. In all results presented in this paper we assume
that center of mass energy of the proton-proton system is $\sqrt
s=14$ TeV.

\section{Sensitivity to anomalous couplings}

We have estimated 95\% confidence level (C.L.) bounds using
one-parameter $\chi^2$ test without a systematic error. The $\chi^2$
function is given by
\begin{eqnarray}
\chi^{2}=\left(\frac{\sigma_{SM}-\sigma_{AN}}{\sigma_{SM} \,\,
\delta}\right)^{2}
\end{eqnarray}
where $\sigma_{AN}$ is the cross section containing new physics
effects and $\delta=\frac{1}{\sqrt{N}}$ is the statistical error.
The number of events is given by $N=\sigma_{SM}L_{int}$ where
$L_{int}$ is the integrated luminosity. ATLAS and CMS have central
detectors with a pseudorapidity coverage $|\eta|<2.5$. Therefore we
place a cut of $|\eta|<2.5$ for final quarks from subprocess $\gamma
q \to \nu \bar \nu q$.

We observe from Eq.(\ref{amplitude1}) that cross section including
$\nu \bar \nu \gamma $ coupling depends on the couplings of the
form; $\sum_{i,j,m,n}\mu_{im}\mu_{mj}\mu_{in}\mu_{nj}$,
$\sum_{i,j}\mu_{ij}^2$ and $\sum_{i,m}\mu_{im}\mu_{mi}$. It receives
contributions from all of the matrix elements
$\mu_{kl},\;(k,l=1,2,3)$. As we have mentioned in the previous
section, some of the matrix elements are strictly constrained by the
experiments. As far as we know the least constrained element is
$\mu_{\tau \tau}$. The bound on this element is $3.9\times10^{-7}
\mu_B$ \cite{Schwienhorst:2001sj}. This bound is at least 3 orders
of magnitude weaker than the bounds on other matrix elements
\cite{Davidson:2005cs}. Therefore in the numerical calculations we
can neglect these strictly constrained elements and assume that only
$\mu_{\tau \tau}$ is nonvanishing. In tables \ref{tab1}, \ref{tab2}
and \ref{tab3} we show 95\% C.L. upper bounds of the couplings
$\mu_{\tau \tau}$, $\alpha_1^2$ and $\alpha_2^2$ for three different
forward detector acceptances. We see from these tables that our
limits on $\mu_{\tau \tau}$ are worse than the current experimental
limit. Our most sensitive limit is approximately 5 times weaker than
the DONUT bound \cite{Schwienhorst:2001sj}. On the other hand we see
from the tables that our bounds on $\alpha_1^2$ and $\alpha_2^2$ are
approximately at the order of $10^{-17}-10^{-18}$. It is more than 8
orders of magnitude more restrictive than the LEP bound. Forward
detector acceptance of $0.1 <\xi < 0.5$ provides more restrictive
bounds on both $\mu_{\tau \tau}$, $\alpha_1^2$ and $\alpha_2^2$
couplings with respect to the $0.0015 <\xi < 0.15$ and $0.0015 <\xi
< 0.5$ cases. This originates from the fact that although the cross
sections in $0.1 <\xi < 0.5$ case are small, deviations of the
anomalous cross sections from the standard model value are
considerably large compare to $0.0015 <\xi < 0.15$ and $0.0015 <\xi
< 0.5$ cases.

\section{Conclusions}

Forward physics program of ATLAS and CMS collaborations provides an
enhancement of the physics studied at the LHC. It allows to study
photon-photon and photon-proton interactions at energies higher than
at any existing collider. LHC as a photon-photon or photon-proton
collider presents an ideal venue to probe neutrino electromagnetic
properties at a high energy.

We have investigated the potential of $p p\to p\gamma p\to p \nu
\bar \nu q X$ reaction at the LHC to probe neutrino-photon and
neutrino-two photon couplings. We show that this reaction has a
great potential to probe neutrino-two photon couplings. It improves
the sensitivity limits by up to a factor of $10^9$ with respect to
LEP limits. Our limits are also better than the limits obtained in
reactions $pp\to p\gamma\gamma p\to p\nu \bar {\nu} p$ and $pp\to
p\gamma\gamma p\to p\nu \bar {\nu}Z p$ at the LHC \cite{lhc10}. On
the other hand, our limits on neutrino-photon coupling $\mu_{\tau
\tau}$ are approximately an order of magnitude worse than the
current experimental bound.

\appendix*
\section{Analytical expressions for $G_1$ and $G_2$}

\begin{figure}
\includegraphics[scale=1]{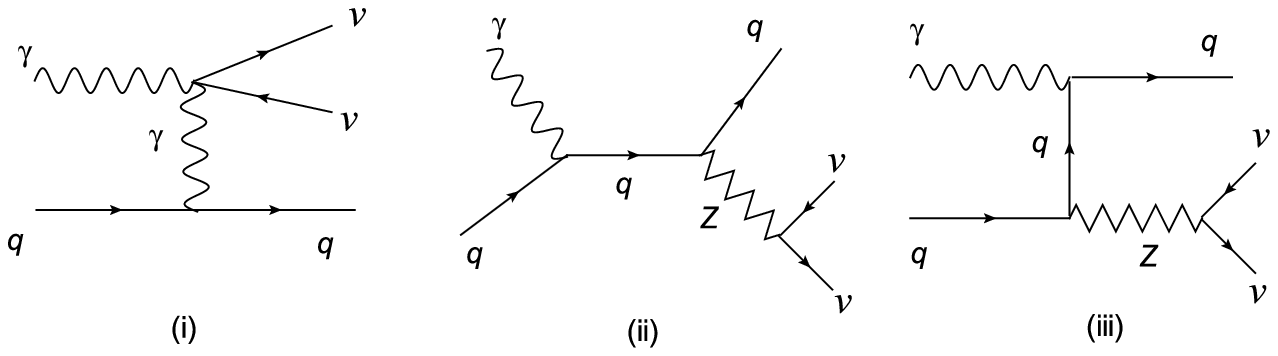}
\caption{Tree-level Feynman diagrams for the subprocess $\gamma q
\to \nu \bar \nu q$ ($q=u,d,c,s,b,\bar u,\bar d, \bar c, \bar s,
\bar b$) in the presence of non-standard $\nu \bar \nu \gamma
\gamma$ coupling.\label{fig3}}
\end{figure}

\begin{figure}
\includegraphics[scale=1]{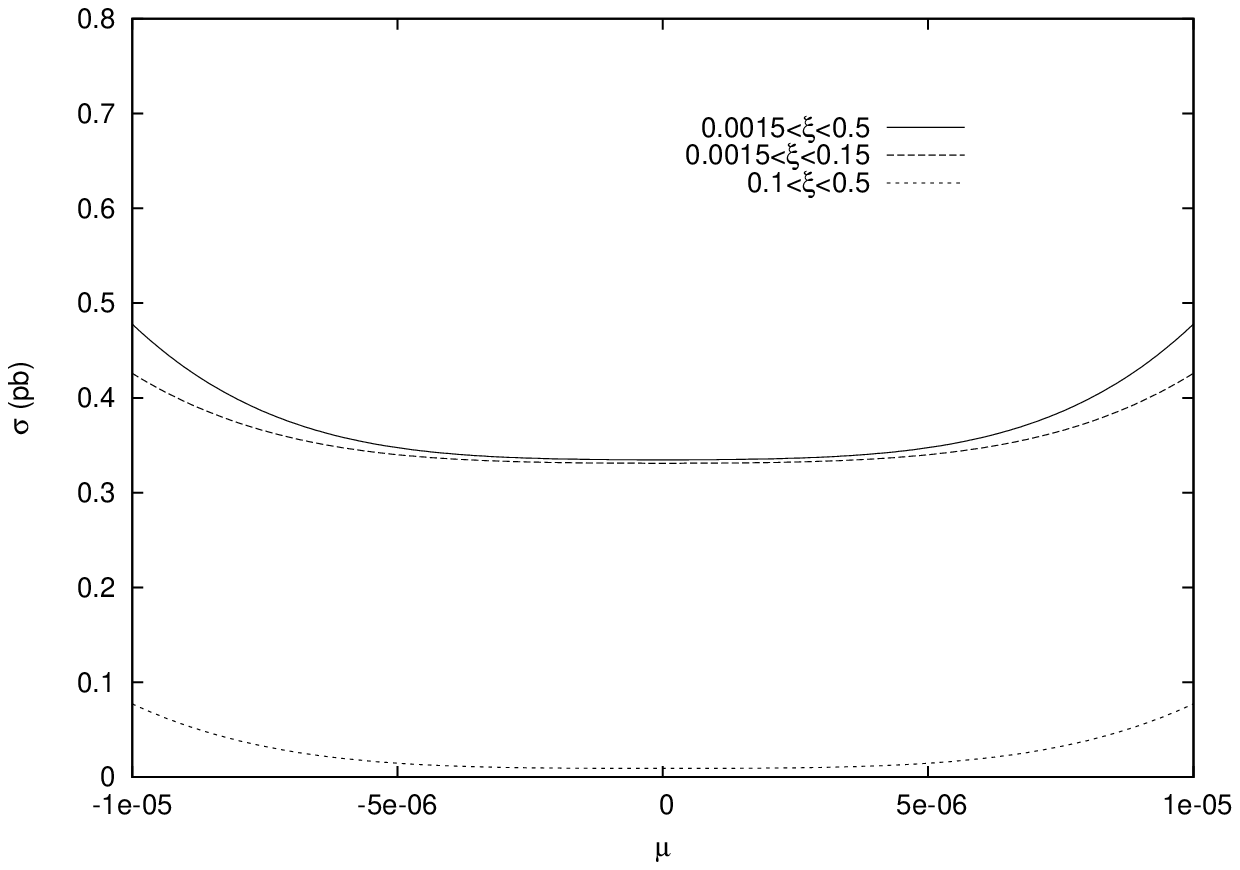}
\caption{The integrated total cross section of the process $p p\to
p\gamma p\to p \nu \bar \nu q X$ as a function of anomalous coupling
$\mu$ for three different forward detector acceptances stated in the
figure. The center-of-mass energy of the proton-proton system is
taken to be $\sqrt s=14$ TeV.\label{fig4}}
\end{figure}

\begin{figure}
\includegraphics[scale=1]{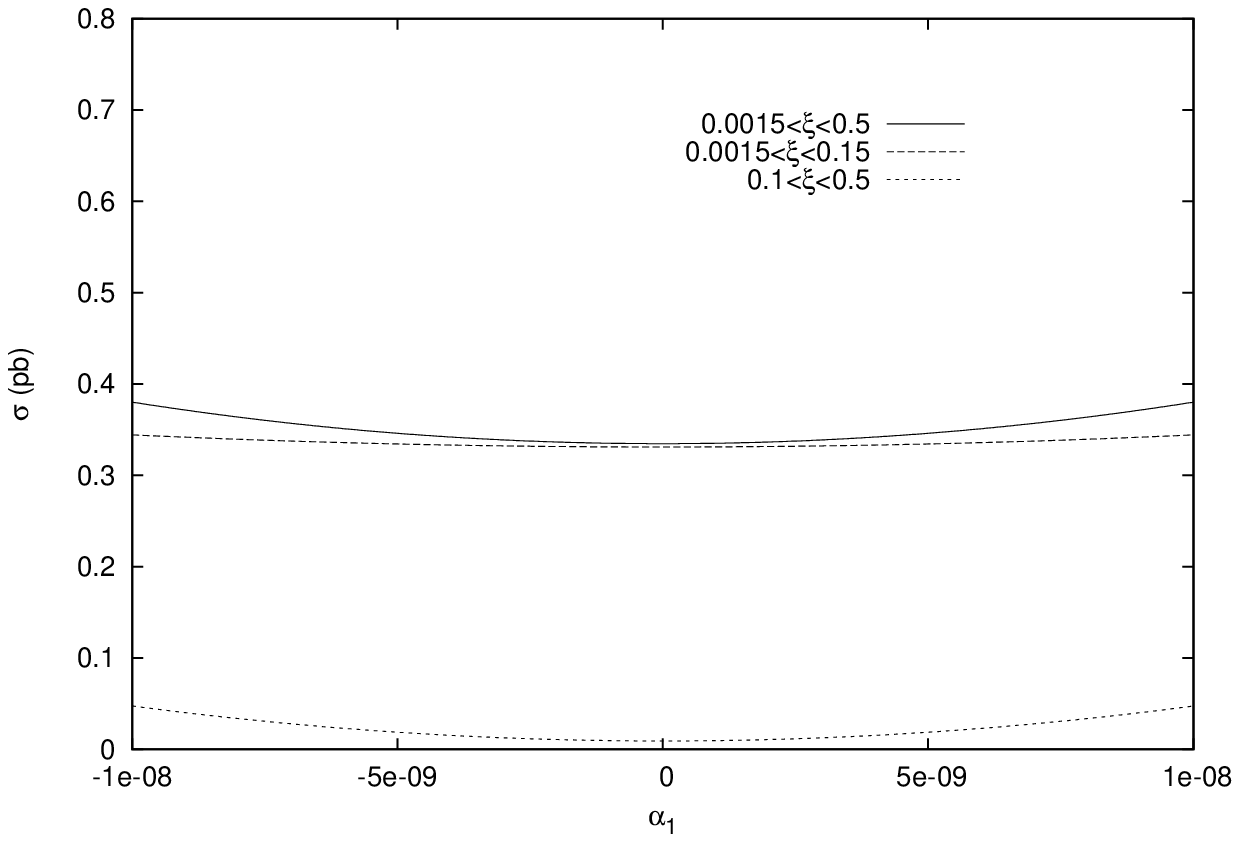}
\caption{The integrated total cross section of the process $p p\to
p\gamma p\to p \nu \bar \nu q X$ as a function of anomalous coupling
$\alpha_1$ for three different forward detector acceptances stated
in the figure. The center-of-mass energy of the proton-proton system
is taken to be $\sqrt s=14$ TeV.\label{fig5}}
\end{figure}

In case of effective interaction $\nu \bar\nu \gamma \gamma$ the
polarization summed amplitude square for the subprocess $\gamma q
\to \nu_i \bar \nu_j q$ $(q=u,d,c,s,b)$ is given by
\begin{eqnarray}
\label{AM1}
\langle|M_1|^2\rangle=&&-\frac{16Q^2g_e^2}{q_1^4}\left[\frac{|\alpha^{ij}_{R1}|^2+|\alpha^{ij}_{L1}|^2
+|\alpha^{ij}_{R2}|^2+|\alpha^{ij}_{L2}|^2}{\Lambda^6}\right]\left\{2m_q^2((p_1\cdot
p_2)^2-p_1\cdot p_5\;p_1\cdot p_2\right. \nonumber \\&&
\left.+(p_1\cdot p_5)^2)-((p_1\cdot p_2)^2+(p_1\cdot p_5)^2)p_2\cdot
p_5\right\}p_3\cdot p_4
\end{eqnarray}
\begin{eqnarray}
\label{AM1213} \langle M_1^\dagger M_2+M_2^\dagger
M_1\rangle=0\;,\;\;\;\;\;\langle M_1^\dagger M_3+M_3^\dagger
M_1\rangle=0
\end{eqnarray}
\begin{eqnarray}
\label{ASM1}
\langle|M_2|^2\rangle=&&\frac{4g_Z^4 Q^2
g_e^2}{(q_2^2-m_q^2)^2(q_3^2-m_Z^2)^2}\left\{-(c^q_A-c^q_V)^2p_4\cdot
p_5\left(m_q^2p_2\cdot p_3+p_1\cdot p_3(m_q^2-p_1\cdot p_2)\right)
\right.\nonumber\\ &&\left.-m_q^2({c^q_A}^2-{c^q_V}^2)
(m_q^2+p_1\cdot p_2)p_3\cdot p_4-(c^q_A+c^q_V)^2p_3\cdot p_5
\left(m_q^2p_2\cdot p_4\right.\right.\nonumber \\&& \left. \left.
+(m_q^2-p_1\cdot p_2)p_1\cdot p_4\right)\right\}\delta_{ij}\\
\label{ASM2} \langle|M_3|^2\rangle=&&\frac{4g_Z^4 Q^2
g_e^2}{(q_4^2-m_q^2)^2(q_3^2-m_Z^2)^2}\left\{
(c^q_A-c^q_V)^2(m_q^2+p_1\cdot p_5)p_1\cdot p_4 \; p_2\cdot
p_3+(c^q_A+c^q_V)^2(m_q^2+p_1\cdot p_5)\right.\nonumber \\&& \left.
\times p_1\cdot p_3\; p_2\cdot p_4-m_q^2\left((c^q_A-c^q_V)^2
p_2\cdot p_3 \; p_4\cdot
p_5+(c^q_A+c^q_V)((c^q_A-c^q_V)(m_q^2-p_1\cdot p_5)p_3\cdot
p_4\right.\right. \nonumber \\&& \left. \left.+(c^q_A+c^q_V)p_2\cdot
p_4\; p_3\cdot p_5)\right)\right\}\delta_{ij}
\end{eqnarray}
\begin{eqnarray}
\label{ASM3} \langle M_2^\dagger M_3+M_3^\dagger
M_2\rangle=&&\frac{4g_Z^4 Q^2
g_e^2}{(q_2^2-m_q^2)(q_4^2-m_q^2)(q_3^2-m_Z^2)^2}\left\{{(c^q_A)}^2\;
p_1\cdot p_4 \;p_2\cdot p_3\; p_2\cdot p_5\right. \nonumber
\\&& \left. +m_q^2\;{(c^q_A)}^2\;p_1\cdot p_2\;p_3\cdot p_4-2\;m_q^2\;
{(c^q_A)}^2p_2\cdot p_5\;p_3\cdot p_4\right. \nonumber \\&&
\left.+{(c^q_A)}^2p_1\cdot p_2\;p_2\cdot p_4\;p_3\cdot
p_5-2\;{(c^q_A)}^2p_2\cdot p_4\;p_2\cdot p_5\;p_3\cdot p_5\;
\right.\nonumber \\&& \left.-2\;c^q_V\;c^q_A\;p_1\cdot p_4\;p_2\cdot
p_3\;p_2\cdot p_5+2\;c^q_V\;c^q_A\;p_1\cdot p_2\;p_2\cdot
p_4\;p_3\cdot p_5\;\right. \nonumber \\&&
\left.-2\;c^q_V\;c^q_A\;p_1\cdot p_4\;p_2\cdot p_5\;p_3\cdot
p_5\;-4\;c^q_V\;c^q_A\;p_2\cdot p_4\;p_2\cdot p_5\;p_3\cdot
p_5\;\right. \nonumber \\&& \left.+(c^q_V)^2\;p_1\cdot p_4\;p_2\cdot
p_3\;p_2\cdot p_5-(c^q_V)^2m_q^2\;p_1\cdot p_2\;p_3\cdot p_4 \right.
\nonumber \\&&\left.+2(c^q_V)^2m_q^2\;p_2\cdot p_5\;p_3\cdot
p_4\;+(c^q_V)^2p_1\cdot p_2\;p_2\cdot p_4\;p_3\cdot p_5\;\right.
\nonumber \\&&\left.-(c^q_V)^2p_1\cdot p_4\;p_2\cdot p_5\;p_3\cdot
p_5\;-2(c^q_V)^2p_2\cdot p_4\;p_2\cdot p_5\;p_3\cdot p_5\;\right.
\nonumber \\&& \left.+\left(p_2\cdot p_3\;(p_1\cdot p_2-2p_2\cdot
p_5)(c^q_A-c^q_V)^2+2({c^q_A}^2+{c^q_V}^2)p_1\cdot p_2\;p_3\cdot
p_5\right)\right. \nonumber \\&& \left.\times p_4\cdot p_5+p_1\cdot
p_3\left(p_2\cdot p_5((c^q_A+c^q_V)^2p_2\cdot
p_4-(c^q_A-c^q_V)^2p_4\cdot p_5)\right.\right. \nonumber
\\&&\left.\left. -2m_q^2({c^q_A}^2-{c^q_V}^2)p_1\cdot
p_4\right)+p_1\cdot
p_5\left((c^q_A+c^q_V)(-(c^q_A-c^q_V)m_q^2p_3\cdot
p_4\right.\right.\nonumber
\\&&\left.\left.-(c^q_A+c^q_V)p_2\cdot p_4\;p_3\cdot
p_5)+p_2\cdot p_3(-p_4\cdot p_5(c^q_A-c^q_V)^2\right.\right.
\nonumber
\\&&\left.\left.-2({c^q_A}^2+{c^q_V}^2)p_2\cdot
p_4)\right)-{(c^q_A)}^2p_1\cdot p_4\;p_2\cdot p_5\;p_3\cdot
p_5\;\right\}\delta_{ij}
\end{eqnarray}
\begin{eqnarray}
g_Z=\frac{g_e}{\sin\theta_W
\cos\theta_W},\;\;\;\;g_e=\sqrt{4\pi\alpha}\nonumber\\
Q=\frac{2}{3},\;\;c^q_A=\frac{1}{2},\;\;c^q_V=\frac{1}{2}-\frac{4}{3}\sin^2\theta_W \;\;\;(q=u,c,t)\nonumber\\
Q=-\frac{1}{3},\;\;c^q_A=-\frac{1}{2},\;\;c^q_V=-\frac{1}{2}+\frac{2}{3}\sin^2\theta_W \;\;\;(q=d,s,b)\nonumber\\
\end{eqnarray}
where $M_1,M_2$ and $M_3$ are the amplitudes of the Feynman diagrams
in Fig.\ref{fig3}. $p_1$ and $p_2$ are the momenta of incoming
photon and quark and $p_3$, $p_4$ and $p_5$ are the momenta of
outgoing neutrino, anti-neutrino and quark respectively. Propagator
momenta are defined by
$q_1=p_2-p_5,\;q_2=p_1+p_2,\;q_3=p_3+p_4,\;q_4=p_5-p_1$. Therefore
$G_1$ and $G_2$ functions in Eq.(\ref{amplitude2}) are defined
through the following equations:
\begin{eqnarray}
(\alpha_1^2+\alpha_2^2)G_1=&&\sum_{i,j}\langle|M_1|^2\rangle\\
G_2=&&3\langle|M_2+M_3|^2\rangle
\end{eqnarray}
In the case of anti-quarks (subprocess $\gamma \bar q \to \nu_i \bar
\nu_j \bar q$) SM amplitudes (\ref{ASM1}), (\ref{ASM2}) and
(\ref{ASM3}) are modified but non-standard contributions (\ref{AM1})
and (\ref{AM1213}) remain unchanged.

\newpage

\begin{table}
\caption{95\% C.L. upper bounds of the couplings $\mu_{\tau \tau}$,
$\alpha_1^2$ and $\alpha_2^2$ for the process $p p\to p\gamma p\to p
\nu \bar \nu q X$. We consider various values of the integrated LHC
luminosities. Forward detector acceptance is $0.0015<\xi<0.5$.
Limits of $\mu_{\tau \tau}$ is given in units of Bohr magneton and
$\Lambda$ is taken to be 1 GeV for limits of $\alpha_1^2$ and
$\alpha_2^2$. \label{tab1}}
\begin{ruledtabular}
\begin{tabular}{cccccc}
Luminosity:&10$fb^{-1}$& 30$fb^{-1}$ &50$fb^{-1}$ &100$fb^{-1}$&200$fb^{-1}$ \\
\hline
$\mu_{\tau \tau}$ &$4.55\times10^{-6}$ &3.79$\times10^{-6}$ &3.47$\times10^{-6}$ &3.06$\times10^{-6}$ &2.68$\times10^{-6}$ \\
$\alpha_1^2$&2.10$\times10^{-17}$ &1.21$\times10^{-17}$ &9.37$\times10^{-18}$ &6.63$\times10^{-18}$ &4.69$\times10^{-18}$   \\
$\alpha_2^2$&2.10$\times10^{-17}$ &1.21$\times10^{-17}$ &9.37$\times10^{-18}$ &6.63$\times10^{-18}$ &4.69$\times10^{-18}$   \\
\end{tabular}
\end{ruledtabular}
\end{table}

\begin{table}
\caption{The same as table \ref{tab1} but for $0.0015<\xi<0.15$.
\label{tab2}}
\begin{ruledtabular}
\begin{tabular}{cccccc}
Luminosity:&10$fb^{-1}$& 30$fb^{-1}$ &50$fb^{-1}$ &100$fb^{-1}$&200$fb^{-1}$ \\
\hline
$\mu_{\tau \tau}$ &5.08$\times10^{-6}$ &4.23$\times10^{-6}$ &3.87$\times10^{-6}$ &3.41$\times10^{-6}$ &2.99$\times10^{-6}$ \\
$\alpha_1^2$&7.16$\times10^{-17}$ &4.13$\times10^{-17}$ &3.20$\times10^{-17}$ &2.27$\times10^{-17}$ &1.60$\times10^{-17}$   \\
$\alpha_2^2$&7.16$\times10^{-17}$ &4.13$\times10^{-17}$ &3.20$\times10^{-17}$ &2.27$\times10^{-17}$ &1.60$\times10^{-17}$   \\
\end{tabular}
\end{ruledtabular}
\end{table}

\begin{table}
\caption{The same as table \ref{tab1} but for $0.1<\xi<0.5$.
\label{tab3}}
\begin{ruledtabular}
\begin{tabular}{cccccc}
Luminosity:&10$fb^{-1}$& 30$fb^{-1}$ &50$fb^{-1}$ &100$fb^{-1}$&200$fb^{-1}$ \\
\hline
$\mu_{\tau \tau}$ &3.40$\times10^{-6}$ &2.82$\times10^{-6}$ &2.58$\times10^{-6}$ &2.27$\times10^{-6}$ &1.98$\times10^{-6}$ \\
$\alpha_1^2$&4.10$\times10^{-18}$ &2.37$\times10^{-18}$ &1.84$\times10^{-18}$ &1.30$\times10^{-18}$ &9.17$\times10^{-19}$   \\
$\alpha_2^2$&4.10$\times10^{-18}$ &2.37$\times10^{-18}$ &1.84$\times10^{-18}$ &1.30$\times10^{-18}$ &9.17$\times10^{-19}$   \\
\end{tabular}
\end{ruledtabular}
\end{table}

\end{document}